\documentclass[usenatbib,letters]{mn2e}
\usepackage{natbib,amssymb,amsmath,times,graphicx,url}

\title[Reynolds numbers in SPH turbulence]{Resolving high Reynolds numbers in SPH simulations of subsonic turbulence}

\author[Price]{Daniel J. Price\thanks{daniel.price@monash.edu}\\
Monash Centre for Astrophysics (MoCA) and School of Mathematical Sciences, Monash University, Vic 3800, Australia \\
}
\date{Submitted: 24th October 2011, Revised: 3rd Nov 2011, Accepted: 5th Nov 2011}
\pagerange{L\pageref{firstpage}--L\pageref{lastpage}} \pubyear{2011}

\begin{document}
\label{firstpage}
\bibliographystyle{mn2e}
\maketitle

\begin{abstract}
Accounting for the Reynolds number is critical in numerical simulations of turbulence, particularly for subsonic flow. For Smoothed Particle Hydrodynamics (SPH) with constant artificial viscosity coefficient $\alpha$, it is shown that the effective Reynolds number in the absence of explicit physical viscosity terms scales linearly with the Mach number --- compared to mesh schemes, where the effective Reynolds number is largely independent of the flow velocity. As a result, SPH simulations with $\alpha=1$ will have low Reynolds numbers in the subsonic regime compared to mesh codes, which may be insufficient to resolve turbulent flow. This explains the failure of \citet[][arXiv:1109.4413v1]{bs11} to find agreement between the moving-mesh code \textsc{arepo} and the \textsc{gadget} SPH code on simulations of driven, subsonic ($v\sim 0.3 c_{\rm s}$) turbulence appropriate to the intergalactic/intracluster medium, where it was alleged that SPH is somehow fundamentally incapable of producing a Kolmogorov-like turbulent cascade.  We show that turbulent flow with a Kolmogorov spectrum can be easily recovered by employing standard methods for reducing $\alpha$ away from shocks.
\end{abstract}


\section{Introduction}
\label{sec:intro}
Turbulence in astrophysics is of key importance for the interstellar (ISM), intracluster (ICM) and intergalactic medium (IGM). Compressible, hydrodynamic turbulence is characterised by two dimensionless parameters, the Mach number $\mathcal{M} \equiv V/c_{\rm s}$, and the Reynolds number \citep{stokes1851,reynolds1883}
\begin{equation}
\mathcal{R}_{\rm e} \equiv \frac{V L}{\nu},
\end{equation}
where $V$ is the flow velocity, $L$ is a typical length scale, $\nu$ is the viscosity of the fluid and $c_{\rm s}$ is the sound speed. Physically, these parameters estimate the relative importance of each of the terms in the Navier-Stokes equations --- the Mach number specifies the ratio of the inertial term, $({\bf v}\cdot\nabla){\bf v}$, to the pressure term, $\nabla P/\rho$, while the Reynolds number specifies the ratio of the inertial term to the viscous dissipation term, $\nu \nabla^{2} {\bf v}$. Mathematically, these two parameters --- along with the boundary conditions and driving --- entirely characterise the flow.

 Given the importance of turbulence in theoretical models, it is crucial that agreement can be found between codes used for simulations of the ISM and ICM/IGM. Several comparison projects have been published recently comparing simulations of both decaying \citep{kitsionasetal09} and driven \citep{pf10} supersonic turbulence relevant to molecular clouds. However, fewer calculations appropriate to the ICM or IGM have been performed. In a recent preprint \citet[][arXiv:1109.4413v1]{bs11} have set out to extend the high Mach number comparisons to the mildly compressible, driven, subsonic turbulence thought to be appropriate to the ICM and IGM. In this case the motions are comparable to or smaller than the sound speed, turbulent motions are dissipated by viscosity, and the flow is mainly characterised by the Reynolds number, similar to turbulence in the laboratory. In particular, it is well known from laboratory studies that the transition from laminar flow to fully developed turbulence only occurs once a critical Reynolds number is reached --- for example for Poiseuille flow (water flowing in a pipe) this is observed for $\mathcal{R}_{\rm e} \gtrsim 2000$ \citep[e.g.][]{reynolds1895}.
 
Since at low Mach number the Reynolds number controls not only the transition to turbulence but also the character of such turbulence (e.g. the extent of the inertial range) it is critical to specify, or at least estimate, the Reynolds number employed in numerical simulations of turbulence in order to compare with the physical Reynolds numbers in the problems of interest. For the ISM, the physical Reynolds numbers are high (e.g. \citealt{es04} estimate $\mathcal{R}_{\rm e}\sim 10^{5}$--$10^{7}$ for the cold ISM) so the approach adopted has been to fix the Mach number and try to reach high numerical Reynolds numbers by minimising numerical dissipation away from shocks. Estimates for $\mathcal{R}_{\rm e}$ in the ICM/IGM are more difficult. \citet{bl07} estimate $\mathcal{R}_{\rm e}\sim 52$ but point out that extremely high values $\gtrsim 10^{10}$ are also possible in the presence of magnetic fields, which change not only the viscosity but also introduce new scales into the problem \citep[see][]{lazarian06a}.  This means that a range of studies are necessary, either by explicitly adding physical viscosity terms to reach a desired (low) $\mathcal{R}_{\rm e}$, or by seeking to minimise numerical viscosity to reach very high $\mathcal{R}_{\rm e}$. A minimum condition for production of turbulent flow in both simulations and reality is that the Reynolds numbers should at least be high enough for the development of turbulence (i.e. $\mathcal{R}_{\rm e} \gtrsim 10^{3}$).

 In their study, \citet{bs11} compare the moving-mesh code \textsc{arepo} (run in both moving and fixed-grid modes) and the Smoothed Particle Hydrodynamics (SPH) code \textsc{gadget-3}. 
The main conclusion of the paper is that, despite good results in the supersonic regime \citep[in agreement with][]{pf10}, ``the widely employed standard formulation of SPH fails quite badly in the subsonic regime'', because of a failure to build up a ``Kolmogorov-like turbulent cascade'' which is apparent in the moving and fixed mesh calculations at the resolutions employed and is expected on theoretical grounds \citep{kolmogorov41} to occur at high Reynolds number. \citet{bs11} attribute the disagreement between the codes to ``large errors in SPH's gradient estimate and the associated subsonic velocity noise'', producing ``essentially unphysical results in the subsonic regime''. This ``casts doubt on the reliability of SPH for simulations of cosmic structure formation''. However, the Reynolds number is neither fixed nor estimated in any of the calculations.

    In this Letter show how the Reynolds numbers in SPH turbulence calculations can be determined (Sec.~\ref{sec:reynolds}). This suggests that in fact the main issue in the SPH calculations employed by \citet{bs11} is that they employ constant --- and thus large --- viscosity parameters for their simulations, which in turn leads to low Reynolds numbers at subsonic velocities, explaining the failure to produce a Kolmogorov-like spectrum. To demonstrate this, we have rerun their calculations (Sec.~\ref{sec:av}), adopting the standard SPH viscosity switch of \citet{mm97} which increases the effective Reynolds number by roughly an order of magnitude and correspondingly leads to results in much better agreement with the grid-based simulations shown in their preprint and with Kolmogorov's scaling relations.
The results are discussed and summarised in Sec.~\ref{sec:discussion}.

\section{Reynolds numbers in SPH turbulence calculations}
\label{sec:reynolds}
 An advantage of SPH, being a Hamiltonian method, is that the only dissipation present in the system is that which is explicitly added. This means that it is straightforward to derive Reynolds numbers for SPH calculations, since any terms added can be directly translated into their physical equivalents. The standard SPH artificial viscosity term in 3D corresponds to Navier-Stokes viscosity terms with shear ($\nu$) and bulk ($\zeta$) viscosity parameters given by \citep[c.f.][]{lp10,price12}
\begin{equation}
\nu \approx \frac{1}{10} \alpha v_{\rm sig} h;  \hspace{1cm} \zeta \approx \frac{1}{6} \alpha v_{\rm sig} h,
\end{equation}
where $h$ is the smoothing length and $\alpha$ is the SPH artificial viscosity parameter (Note that in 2D the parameters are $1/8$ and $5/24$, respectively, c.f. \citealt{murray96,monaghan05,price12}). For low Mach number calculations we can expect that the $\nabla (\nabla\cdot {\bf v})$ terms represent only a small contribution to the overall dissipation rate, such that the dissipation is mainly controlled by the shear viscosity term. Furthermore the maximum signal velocity $v_{\rm sig} \approx c_{\rm s}$ at low Mach number, such that the effective Reynolds number can be easily computed according to
\begin{equation}
\mathcal{R}_{\rm e} \equiv \frac{V L}{\nu} = \frac{10}{\alpha} \mathcal{M} \frac{L}{h},
\end{equation}
where $\mathcal{M}$ is the Mach number. Note that, with $\mathcal{M}$ fixed, the Reynolds number is determined entirely by two parameters: the value of $\alpha$ and the numerical resolution $h/L$. For simulations in a periodic box using the parameters employed by \citet{bs11} we have
\begin{equation}
\mathcal{R}_{\rm e} = 2.4 n_{\rm x} \left( \frac{\mathcal{M}}{0.3}\right) \left( \frac{\alpha}{1.0}\right)^{-1} \left(\frac{N_{\rm ngb}}{64}\right)^{-1/3},
\label{eq:Re}
\end{equation}
where $n_{\rm x}$ is the number of particles along the box length $L$ and $N_{\rm ngb}$ is the neighbour number parameter in \textsc{gadget-3}, which corresponds to
\begin{equation}
N_{\rm ngb, 3D} = \frac{4}{3} \pi (\eta R)^{3},
\end{equation}
where $R$ is the kernel truncation radius in units of h (i.e. $R=2$ for the cubic spline kernel) and $\eta$ is the usual parameter specifying the smoothing length in units of the particle spacing ($h = \eta [m/\rho]^{1/n_{\rm dim}}$). Referring to the neighbour number is in general misleading since it is not independent of the resolution length $h$ meaning that increasing $N_{\rm ngb}$ also corresponds to increasing $h$ (see discussion in \citealt{price12}). In this Letter we use $\eta = 1.2$ corresponding to $\sim 58$ neighbours in a uniform particle distribution.

 From Eq.~\ref{eq:Re} it is easy to see why simulations are difficult at low Mach number ($\mathcal{M} \lesssim 0.5$) in SPH with constant viscosity parameters. For the resolutions employed by \citet{bs11}, namely $n_{x} = 64$, $128$ and $256$ the Reynolds numbers are given by $\mathcal{R}_{\rm e} = 154$, $\mathcal{R}_{\rm e} = 307$ and $\mathcal{R}_{\rm e} = 614$ respectively. By comparison, the Reynolds numbers that would be reached even with $\alpha=1$ at Mach 10 would be approximately 30 times higher, giving $\mathcal{R}_{\rm e} \approx 18,000$ at $256^{3}$ and $\mathcal{R}_{\rm e} \approx 37,000$ at $512^{3}$ particles. \citet{pf10} also employed viscosity switches that decrease the viscosity away from shocks by up to an order of magnitude, meaning that they obtain Reynolds numbers of $\sim 10^{5}$ and higher in practice (a plot of the Reynolds number in the \citealt{pf10} calculations is shown in Fig.~2 of \citealt{pf10b}; $\mathcal{R}_{\rm e}$ of up to $10^{6}$ are achieved in the densest regions where the SPH resolution is highest).

\begin{figure*}
   \centering
   \includegraphics[width=\textwidth]{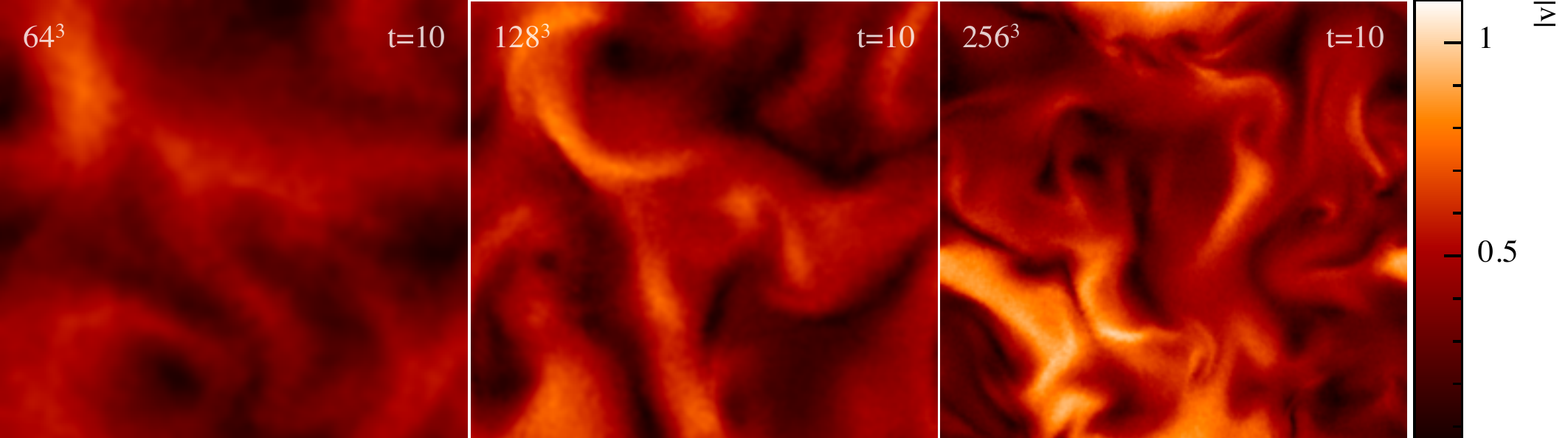} 
   \includegraphics[width=\textwidth]{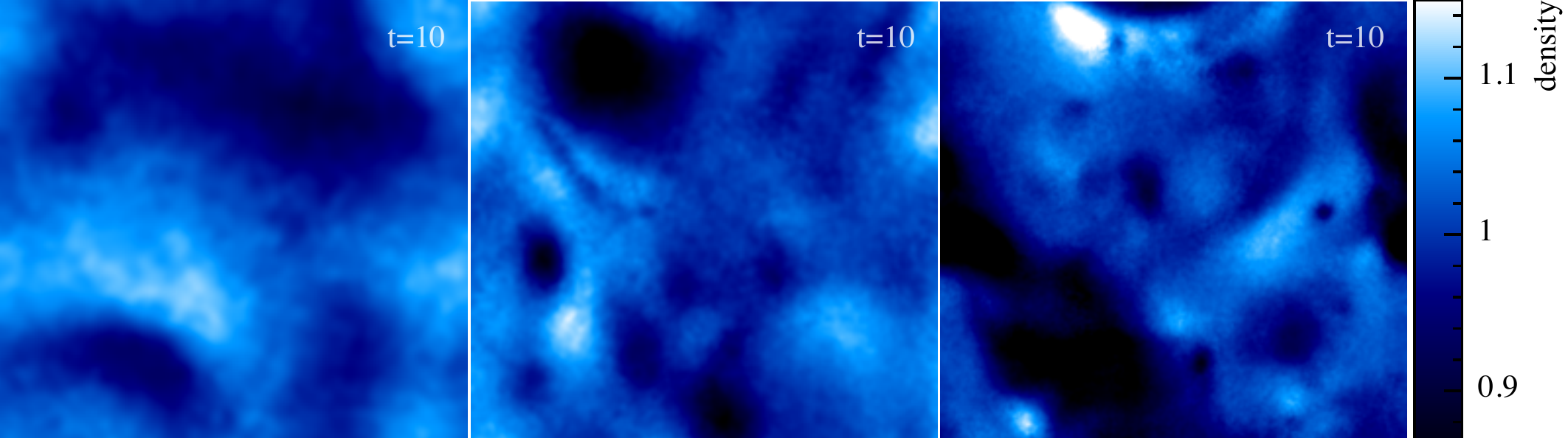} 
   \includegraphics[width=\textwidth]{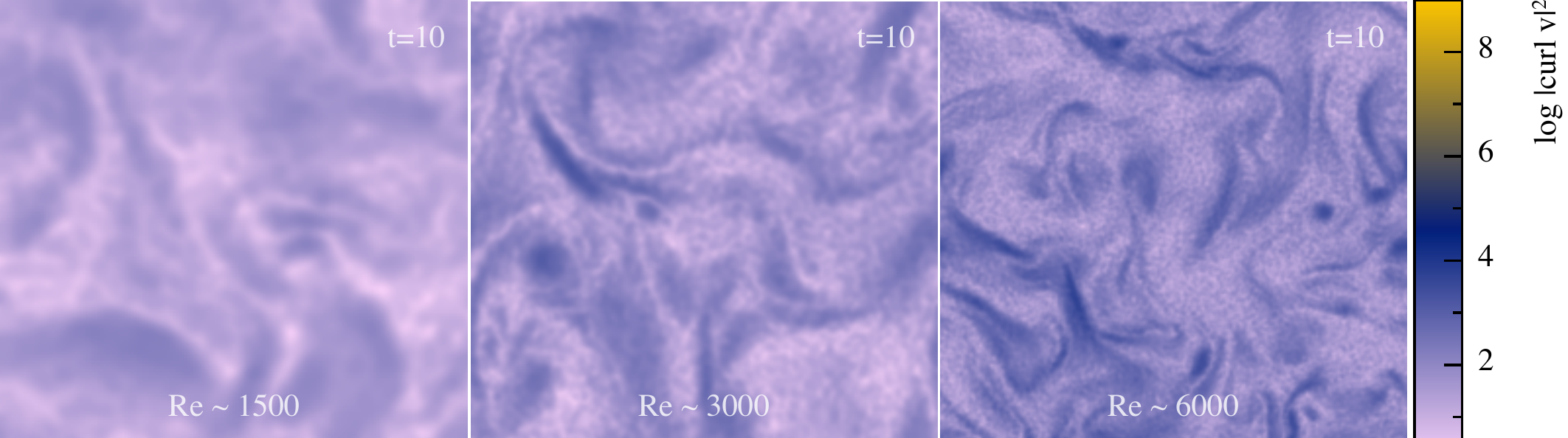} 
   \caption{Cross section slices through the midplane showing magnitude of the velocity field (top row), density (second row) and $\vert \nabla \times {\bf v}\vert^{2}$ (bottom row) at $t=10$ for SPH calculations with the \textsc{phantom} code employing the \citet{mm97} viscosity switch at resolutions of $64^{3}$, $128^{3}$ and $256^{3}$ particles (left to right). The $256^{3}$ calculations may be compared to the corresponding panels in Fig.~3 of \citet{bs11}. Estimates of the effective Reynolds numbers are indicated for each calculation.}
   \label{fig:absv}
\end{figure*}

 Note also that the situation regarding numerical viscosity is very different in SPH compared to grid-based codes. The intrinsic numerical dissipation in a grid-based code is in general linearly proportional to the advection velocity relative to the grid \citep[e.g.][]{robertsonetal10}, whereas in SPH the artificial viscosity term is linearly proportional to the resolution and largely independent of the velocity except at high Mach number where the $\beta$ term comes into play. This means that Reynolds numbers are roughly constant in Eulerian schemes over a range of Mach numbers --- because the increase in V is correspondingly offset by the increase in the numerical viscosity, whereas the Reynolds number in SPH has a linear Mach number dependence. Given that good agreement was found between grid and SPH codes in the power spectra in the \citet{pf10} calculations, it is likely that the Reynolds number achievable in the fixed and moving grid schemes employed in \textsc{arepo} are of similar magnitude (i.e. $\sim 10^{5}$ at $512^{3}$ and slightly lower at $256^{3}$).

\section{Reducing the viscosity in SPH}
\label{sec:av}
\subsection{Standard approaches}
 \citet{bs11} do not use any viscosity switches for their main calculations, despite the fact that most of these switches are at least $\sim 15$ years old and in widespread use. The standard viscosity switch in use is the one proposed by \citet{mm97}, where $\alpha$ is a time-dependent parameter that responds to a source term proportional to $-\nabla\cdot{\bf v}$ (i.e., converging flows) and in the absence of such terms decays to a minimum $\alpha_{\rm min}$, typically set to $0.1$. Already use of this switch would substantially increase the Reynolds number, though even a factor of 10 reduction in $\alpha$ gives only $\mathcal{R}_{\rm e} \approx 6000$ for their ``S3'' calculation which is still a far cry from the $\mathcal{R}_{\rm e} \approx 10^{5}$ achieved by \citet{pf10}. 
  
\begin{figure}
   \centering
   \includegraphics[width=\columnwidth]{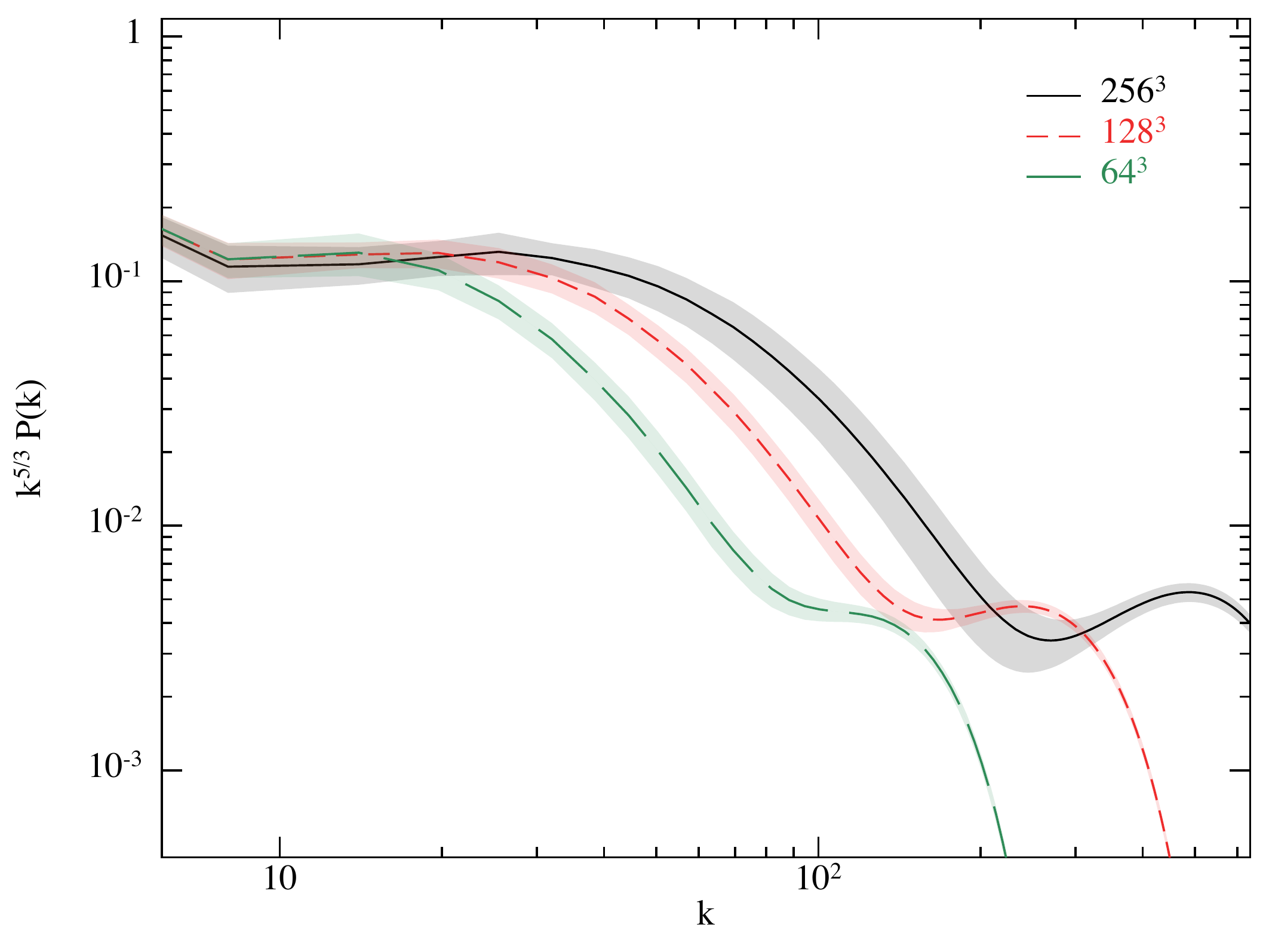}
   \caption{Time-averaged $k^{5/3}$-compensated power spectra from subsonic SPH turbulence calculations using the \citet{mm97} viscosity switch at a resolution of $64^{3}$, $128^{3}$ and $256^{3}$ particles, as indicated, for which the corresponding Reynolds numbers are $\sim 1500$, $3000$ and $6000$, respectively. The shaded regions show the 1$\sigma$ errors from the time-averaging. At the highest Reynolds numbers a Kolmogorov-like $k^{-5/3}$ slope is evident at large scales.}
   \label{fig:pspec}
\end{figure}

 Fig.~\ref{fig:absv} shows the results of a series of 3 calculations performed with the \textsc{phantom} SPH code \citep{pf10,lp10}, employing the same driving routine as described in \citet{pf10} (adapted from \citealt{federrathetal08,federrathetal10}) with the same parameters as the \citet{bs11} calculations\footnote{The SPH version of the driving routine can be made available on request. The parameters used here are $st_{\rm energy} = 0.002$, $st_{\rm decay} = 1.0$, $st_{\rm solweight} = 1.0$, $st_{\rm stirmin}=6.28$, $st_{\rm stirmax} = 18.85$, $st_{\rm dtfreq} = 0.005$, corresponding to the stirring energy, decay timescale, solenoidal driving, minimum and maximum wavenumbers and the frequency with which the stirring is updated. The driving is given a $k^{-5/3}$ wavenumber dependence in the stirring range as described in \citet{bs11} and denoted as $st_{\rm spectform} = 2$ in the input file. The random number generator and seed will however differ from their calculations.} but with the \citet{mm97} switch with $\alpha_{\rm min} = 0.05$, resulting in $\overline{\alpha}\approx 0.1$. The runs are performed in a periodic box  $x,y,z \in [0,1]$, using an isothermal equation of state with a sound speed in code units of unity. Although the Reynolds numbers achieved are evidently still lower than in the grid-based calculations employed by \citet{bs11}, it is clear that already this is a dramatic improvement on the SPH simulations shown in their preprint. The resulting power spectra are shown in Fig.~\ref{fig:pspec}, showing the time-averaged spectrum from 191 snapshots sampled every $\Delta t = 0.1$ between $t=6$ and $t=25$. The $y-$axis shows the power spectrum compensated by $k^{5/3}$ such that a $k^{-5/3}$ spectrum would appear horizontal. Though the spectrum ``turns over'' at relatively small $k$ at low resolution, a clear $k^{-5/3}$ range is apparent in the highest resolution calculations. The resolution dependence of the high $k$ turnover is also consistent with the expected $\mathcal{R}_{\rm e}^{-3/4}$ dependence of the dissipation scale.

\subsection{The state of the art}
  With mean motions that are around $1/3$ of the sound speed and transsonic fluctuations, one cannot simply reduce the SPH artificial viscosity parameters arbitrarily (such as the $\alpha = 0.1$ and $\alpha = 0$ calculations attempted by \citealt{bs11}), since this term is necessary to capture the physical dissipation that occurs due to the non-linear steepening of waves. Such an approach may be adequate for very low Mach number (i.e., incompressible) calculations but it provides no easy answer at $\sim 0.3 c_{\rm s}$. Instead it is clear that to achieve similar results with SPH an improved viscosity switch is necessary in order to both capture non-linear steepening and shocks as well as reducing the viscosity to very low values where it is not needed. The switch proposed recently by \citet{cd10} represents the current state-of-the-art in this regard, essentially a thoroughly enhanced and improved version of the \citet{mm97} approach. In particular, they show that they are able to simulate linear waves for over 50 periods with essentially no numerical dissipation, using the same parameters as would be applied in shock problems. Thus, with an implementation of the \citet{cd10} switch it may be expected that significantly higher Reynolds numbers are achievable in SPH.
  
\section{Discussion}
\label{sec:discussion}
  \citet{bs11} argue that ``large errors in SPH's gradient estimate'' are responsible for the failure to reproduce a Kolmogorov-like turbulent cascade in their SPH calculations. Figure~\ref{fig:pspec} demonstrates that this argument is incorrect, since we are able to obtain a $k^{-5/3}$ spectrum using only standard SPH gradient terms and a very similar SPH neighbour number to that employed in their preprint. However, we find that the appearance of a power-law inertial range in the power spectrum strongly depends on the Reynolds number employed in the calculations, requiring at least $\mathcal{R}_{e} \gtrsim 1500$. This explains the failure to produce a turbulent cascade in their SPH results, since the maximum Reynolds numbers they achieve are $\approx 600$. With the \citet{mm97} viscosity switch employed in this Letter we estimate that we are able to achieve $\mathcal{R}_{e} \approx 6,000$ at $256^{3}$ particles which already brings the SPH results into much better agreement with the grid-based results shown in their work. Indeed, both \citet{dolagetal05} and \citet{valdarnini11} have already pointed out that using this switch could substantially improve SPH simulations of turbulence in galaxy clusters.

 It should be noted that \citet{bs11} do experiment with reduced viscosity parameters in their preprint, using either a fixed $\alpha = 0.1$ or the \citet{balsara95} switch and also a run with zero viscosity (as we have already discussed in Sec.~\ref{sec:av}, it is not clear that one can simply reduce the parameters arbitrarily, so this approach is questionable --- particularly the $\alpha=0$ calculation). Indeed, both the $\alpha = 0.1$ and Balsara-switch calculations show a dramatic improvement in the power spectrum at large scales. The authors dismiss this result because of a corresponding increase in power at $k \gtrsim 100$. However, the power at these scales is low amplitude ($\sim 10^{-4}$) and thus sensitive to all manner of numerical artefacts (e.g. the interpolation procedure as demonstrated in Fig.~4 of their preprint). Indeed, we do not find the upturn in power at large $k$ seen in their results (c.f. Fig.~\ref{fig:pspec}), most likely due to our improved power spectrum estimation --- here computed by interpolating the SPH data to a 3D grid using the kernel and employing a Fast Fourier Transform, rather than the ``nearest neighbour sampling'' procedure employed in their preprint.

Finally, it is important to compare the Reynolds numbers achievable in numerical simulations to the physical Reynolds numbers in the problems of interest (c.f. Sec.~\ref{sec:intro}). For the cold ISM, $\mathcal{R}_{\rm e}\sim 10^{5}$--$10^{7}$ which, though high, is not as high as is often assumed, and is certainly within reach of being resolved with currently achievable resolutions (c.f. Sec.~\ref{sec:reynolds}). This implies at the very least that physical viscosity should be introduced into ISM turbulence simulations in the near future. In the ICM/IGM, $\mathcal{R}_{\rm e}\sim 52$ would seem to imply laminar flow, though very high estimates for $\mathcal{R}_{\rm e}$ ($\gtrsim 10^{10}$) apply in the presence of magnetic fields. Reaching such Reynolds numbers is not presently achievable with any numerical code. However, this may also imply that ultimately it is quite incorrect to try to simulate purely hydrodynamic ICM/IGM turbulence at high Reynolds number without taking into account more detailed physics, such as magnetic fields.

\section{Conclusions}
 In this Letter we have emphasized the importance of accounting for the Reynolds number in numerical turbulence simulations, particularly in the subsonic regime where it is the main parameter controlling not only whether the flow is turbulent but also the character of such turbulence. In particular differences in the Reynolds number can produce strong \emph{physical} differences between calculations, independent of any numerical influences. We have shown that use of viscosity switches is the key to higher Reynolds numbers in SPH turbulence calculations, producing spectra consistent with Kolmogorov theory.
 
\section*{Acknowledgments}
We are very grateful to the referee, Walter Dehnen, both for constructive feedback and prompt reviewing of this Letter. We thank Andreas Bauer and Volker Springel for kindly supplying the driving parameters used for the calculations. We also thank Chris Nixon, Jasmina Lazendic-Galloway, Duncan Galloway, Joe Monaghan, Jules Kajtar, Terry Tricco, Guillaume and Florence Laibe for useful discussion and comment. Figures were made using \textsc{splash} \citep{splashpaper} with the new \textsc{giza} backend. Calculations were run on the Monash Sun Grid, with thanks to Philip Chan and MERC.

\bibliography{sph,turbulence,starformation}

\label{lastpage}
\enddocument